\documentclass[conference]{IEEEtran}
\usepackage{cite}
\usepackage{amsmath,amssymb,amsfonts}
\usepackage{graphicx}
\usepackage{xcolor}
\usepackage{tikz}
\usepackage[margin=1in]{geometry}

\usetikzlibrary{automata, positioning, arrows, calc}

\def\BibTeX{{\rm B\kern-.05em{\sc i\kern-.025em b}\kern-.08em
    T\kern-.1667em\lower.7ex\hbox{E}\kern-.125emX}}

\tikzstyle{block} = [draw, fill=white, rectangle, minimum height=3em, minimum width=4em, thick]
\tikzstyle{sum} = [draw, fill=white, circle, thick]
\tikzstyle{input} = [coordinate]
\tikzstyle{noise} = [coordinate]
\tikzstyle{output} = [coordinate]
\tikzstyle{output1} = [coordinate]
\tikzstyle{disturbance} = [coordinate]
\tikzstyle{pinstyle} = [pin edge={to-,thin,black}]

\newcommand{\elab}[1]{\label{eqn:#1}}
\newcommand{\eqn}[1]{(\ref{eqn:#1})}

\newcommand{\flab}[1]{\label{fig:#1}}
\newcommand{\fig}[1]{Fig.\ref{fig:#1}}

\newcommand{\vo}[1]{\boldsymbol{#1}}
\newcommand{\etal}{\textit{et al.}}

\newcommand{\uvec}[1]{\hat{\boldsymbol{#1}}}

\begin{document}

\title{Robust Cislunar Navigation via LFT-Based $\mathcal{H}_\infty$ Filtering with Bearing-Only Measurements}

\author{Raktim Bhattacharya \\[2mm] \small  Aerospace Engineering, Texas A\&M University \\ College Station, TX 77843-31418, USA \\[2mm] \small Email: raktim@tamu.edu}

\maketitle

\begin{abstract}
This paper develops a robust estimation framework for cislunar navigation that embeds the 
Circular Restricted Three-Body Problem (CR3BP) dynamics and bearing-only optical 
measurements within a Linear Fractional Transformation (LFT) representation. A full-order 
$\mathcal{H}_\infty$ observer is synthesized with explicit $\mathcal{L}_2$ performance bounds. 
The formulation yields a nonlinear estimator that operates directly on the governing equations 
and avoids reliance on local linearizations. Dominant nonlinearities are expressed as 
structured real uncertainties, while measurement fidelity is represented through range-dependent weighting with Earth-Moon distances reconstructed from line-of-sight geometry. The sensing architecture assumes passive star-tracker-class optical instruments, eliminating the need for time-of-flight ranging or precision clocks. Simulations demonstrate bounded estimation errors and smooth position tracking over multiple orbital periods, with the largest deviations observed in the out-of-plane states, consistent with the stiffness of the vertical dynamics and the limitations of angle-only observability. Application to a Near Rectilinear Halo Orbit (NRHO) illustrates that the framework can achieve 
robust onboard navigation with bounded estimation errors with flight-representative sensors.
\end{abstract}


\section{Introduction}
The cislunar domain has emerged as a focal area for exploration, with Near Rectilinear Halo Orbits (NRHOs) identified as strategically advantageous for future missions. NASA's Artemis Gateway is baselined to operate in an NRHO, providing a persistent staging node for lunar surface operations and deep-space activities.

Navigation in cislunar space presents challenges distinct from conventional Earth-orbiting missions. The multi-body gravitational environment induces strongly nonlinear dynamics with pronounced sensitivity to initial conditions, complicating long-horizon state prediction. The absence of Global Navigation Satellite System (GNSS) coverage removes a primary navigation aid available in low Earth orbit. Earth-based tracking suffers from limited geometric diversity and large slant ranges, degrading observability during certain orbital phases. Moreover, periodic lunar occultations (lasting 8-14 hours per orbital period) combined with Deep Space Network scheduling limitations can produce extended communication gaps, during which the spacecraft must rely exclusively on onboard estimation. The weak gravitational regime amplifies the impact of small perturbations—such as solar radiation pressure, higher-order gravitational harmonics, and third-body effects—necessitating high-fidelity force modeling. The requirement for autonomous navigation follows from stringent operational constraints, where missed or mistimed station-keeping maneuvers can drive rapid trajectory divergence and incur propellant penalties, while time-critical activities such as rendezvous or preparations for lunar descent demand uninterrupted onboard navigation independent of communication geometry.

Autonomous navigation in cislunar space, particularly NRHOs, requires sophisticated estimation techniques capable of handling the unique challenges of the cislunar environment. Current autonomous navigation techniques face significant limitations when applied to the NRHO environment. The Extended Kalman Filter (EKF), while computationally efficient and flight-proven \cite{montenbruck2000satellite}, struggles with the strongly nonlinear multi-body dynamics. Linearization errors accumulate rapidly during  communication outages, causing position uncertainties to exceed mission requirements \cite{carpenter2025navigation}. Advanced nonlinear filtering techniques offer improved performance but with trade-offs. The Unscented Kalman Filter (UKF) propagates sigma points through nonlinear dynamics, achieving 20-40\% position accuracy improvements over EKF \cite{julier2004unscented}. Particle filters can handle non-Gaussian uncertainties but require significant computational resources \cite{gordon1993novel}. Consider-covariance filters (or Schmidt-Kalman filter) treat uncertain parameters as random variables in covariance propagation while maintaining fixed state estimates, proving effective for slowly varying uncertainties. Recent work shows a practical application in NRHOs for a hybrid constellation. It demonstrates how crosslinks mitigate NRHO estimation weaknesses \cite{iiyama2021autonomous} but leaves open the challenge of guaranteed robustness. Multiple Model Adaptive Estimation (MMAE) uses filter banks for different operational modes \cite{bar2001estimation}, yet struggles with the rapid mode transitions characteristic of NRHO operations. 

Despite these advances, fundamental challenges remain in maintaining navigation accuracy 
under limited measurement availability and strongly nonlinear dynamics. This motivates the 
development of robust estimation methods that provide guaranteed performance bounds across 
all operational conditions.

This paper presents a robust autonomous navigation framework for cislunar space using LFT models of the nonlinear CR3B dynamics and $\mathcal{H}_\infty$ optimization. The key contributions include:
\begin{itemize}
    \item Development of a Linear Fractional Transformation (LFT) model capturing the nonlinear CR3BP dynamics, perturbation uncertainties, and range-dependent sensor noise.
    \item Design of a robust $\mathcal{H}_\infty$ observer to ensure guaranteed $\mathcal{L}_2$ performance in the presence of bounded system uncertainties
    \item Performance analysis through simulations demonstrating robustness and accuracy under realistic cislunar conditions.
\end{itemize}
This is related to our previous work on two dimensional CR3BP \cite{kumar2025cislunar} navigation in cislunar constellations. This paper addresses the more complex three dimensional navigation problem, demonstrated on NRHO trajectories.

\section{Cislunar Dynamics and Sensing Characteristics}
\subsection{The CR3BP Model}
\begin{figure}[ht!]
\centering
\begin{tikzpicture}[>=latex',thick, scale=0.535, every node/.style={scale=0.535}]
    
\coordinate (O) at (0,0){};
\node[draw,circle,name=E,fill=blue!20] at (-3,0){\Huge $E$};
\node[draw,circle,name=M,fill=gray!10] at (10,0){\large$M$};
\node[draw,circle, name=s,fill=green!30] at (3,3){$s$};

\node[below of=E]{$(-\mu,0,0)$}; \node[above of=E]{mass $m_1$};
\node[below of=M]{$(1-\mu,0,0)$}; \node[above of=M]{mass $m_2$};
\node[above of=s]{$(x,y,z)$};




\draw[->] (E) -- (s) node[midway, above left]{$\vo{r}_{1}$};
\draw[->] (M) -- (s) node[midway, above right]{$\vo{r}_{2}$};
\draw[->] (O) -- (s) node[midway, above, xshift=5mm]{$\vo{r}$};
\draw[-,thin] (E) -- (M) node[midway, above]{};

\draw[->,red,very thick, opacity=0.75] (O) -- +(1,0) node[above]{$x$};
\draw[->,red,very thick, opacity=0.75] (O) -- +(0,1) node[right]{$y$};
\filldraw[fill=red,opacity=0.75] (O) circle (2pt) node[below=2mm,opacity=0.75,color=red]{$O$};
\end{tikzpicture}
\caption{Cislunar Rotating Reference Frame, with $E$ (Earth), $M$ (Moon), and $s$ (Satellite).}
\flab{cislunar}
\end{figure}
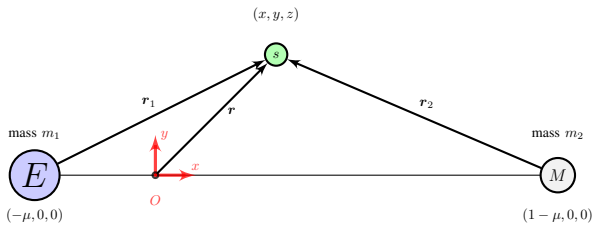
The Circular Restricted Three-Body Problem (CR3BP) provides a simplified yet effective model for studying spacecraft dynamics in the Earth-Moon system. In this framework, we consider two massive primaries (Earth and Moon) orbiting their common barycenter in circular orbits, while a third body (spacecraft) of negligible mass moves under their gravitational influence without affecting their motion. We establish a rotating coordinate frame with origin at the system barycenter. As shown in \fig{cislunar}, the $x$-axis points along the line connecting the two primaries, the $z$-axis is perpendicular to the orbital plane, and the $y$-axis completes the right-handed system. This frame rotates with constant angular velocity $\omega$ equal to the mean motion of the primaries. To simplify analysis, we normalize units such that the distance between primaries is unity, the sum of primary masses is unity, the gravitational constant $G = 1$, the angular velocity $\omega = 1$, and the orbital period of primaries is $2\pi$.

The mass ratio parameter $\mu = m_2/(m_1 + m_2)$ characterizes the system, where $m_1$ and $m_2$ are the masses of the larger (Earth) and smaller (Moon) primary bodies respectively. For the Earth-Moon system, $\mu \approx 0.01215$.

In the normalized rotating frame, the larger primary (Earth) is positioned at $(-\mu, 0, 0)$ and the smaller primary (Moon) at $(1-\mu, 0, 0)$.

The energy-based derivation proceeds from the effective potential,
\begin{equation}
U(x,y,z) = \frac{1}{2}(x^2 + y^2) + \frac{1-\mu}{r_1} + \frac{\mu}{r_2},
\end{equation}
where 
\begin{subequations}
\begin{align}
r_1 &= \sqrt{(x+\mu)^2 + y^2 + z^2} \text{ and }\\
r_2 &= \sqrt{(x-(1-\mu))^2 + y^2 + z^2}
\end{align} 
\elab{r1r2}    
\end{subequations}

are the distances from the third body to each primary. The kinetic energy in the rotating frame is
\begin{equation*}
T = \frac{1}{2}(\dot{x}^2 + \dot{y}^2 + \dot{z}^2).
\end{equation*}

Applying the Lagrangian formulation $\mathcal{L} = T - U$, and using the Euler-Lagrange equations,
\begin{equation*}
\frac{d}{dt}\left(\frac{\partial \mathcal{L}}{\partial \dot{q}_i}\right) - \frac{\partial \mathcal{L}}{\partial q_i} = 0,
\end{equation*}
where $q_i \in \{x,y,z\}$, we obtain the equations of motion
\begin{align}
\ddot{x} - 2\dot{y} = \frac{\partial U}{\partial x}, && \ddot{y} + 2\dot{x} = \frac{\partial U}{\partial y}, && \ddot{z} = \frac{\partial U}{\partial z}.
\elab{cr3bp_dynamics}
\end{align}
The partial derivatives of the effective potential are given by
\begin{align*}
\frac{\partial U}{\partial x} &= x - \frac{(1-\mu)(x+\mu)}{r_1^3} - \frac{\mu(x-(1-\mu))}{r_2^3},\\
\frac{\partial U}{\partial y} &= y - \frac{(1-\mu)y}{r_1^3} - \frac{\mu y}{r_2^3},\\
\frac{\partial U}{\partial z} &= -\frac{(1-\mu)z}{r_1^3} - \frac{\mu z}{r_2^3}.
\end{align*}
The CR3BP admits a conserved quantity, the Jacobi constant
\begin{equation}
C = 2U - (\dot{x}^2 + \dot{y}^2 + \dot{z}^2).
\end{equation}

While the CR3BP provides a foundational framework for NRHO dynamics, several perturbations introduce model uncertainties that must be considered for high-fidelity navigation. The primary sources of uncertainty include solar radiation pressure, third-body gravitational effects (notably from the Sun), higher-order gravitational harmonics of Earth and Moon, maneuver execution errors, and relativistic corrections. Each perturbation contributes to trajectory deviations that can accumulate significantly over the multi-day orbital periods characteristic of NRHOs.

\section{Design of Robust Navigation Observer}
\subsection{Background on LFT Modeling of Nonlinear Systems} 
Linear Fractional Transformation (LFT) modeling provides a systematic framework for representing nonlinear systems within a structured uncertainty paradigm, enabling the application of robust control/estimation synthesis techniques \cite{zhou1998essentials}. As illustrated in \fig{lft_basics}, the fundamental premise is to separate a nonlinear system into two components: a linear time-invariant (LTI) nominal plant and a structured uncertainty block that captures the nonlinear behavior.

\begin{figure}[ht!]
\begin{tikzpicture}[auto, node distance=2.5cm, >=latex', scale=0.7, every node/.style={scale=0.7}]

    \node [input] at (0,0) (d) {};
    \node [block, right of=d, node distance=2cm] (P1) {$\begin{array}{l}\dot{x} = f(x,w,\Delta(t)), \\ y = h(x,w,\Delta(t)) \end{array}$};
    \node [output, right of=P1, thick,node distance=2cm] (y) {};    
    \node [below of=P1, yshift=1.5cm] (label1) {Nonlinear System};
    \draw [->] (d) node[left]{$w$} -- (P1);
    \draw [->] (P1) -- (y) node[right]{$y$};

    \node (LFT) at ($(y)+(1,0)$){$\Longrightarrow$};
    \node [above of=LFT, yshift=-2.0cm] (label3) {LFT};

    \node [block,right of=LFT,node distance=3cm] (P) {$M$};
    \node [below of=P, yshift=1.5cm] (label2) {LFT Representation};
    \node [block, above of=P,node distance=1.5cm] (delta) {$\Delta(t)$};

    \node [input, left of=P, yshift=2mm,node distance=1.5cm] (w1) {};
    \node [input, left of=P, yshift=0mm] (w2) {};    

    \node [output, right of=P, xshift=-.8cm, yshift=2mm, node distance=1.5cm] (z1) {};
    \node [output, right of=P, yshift=0mm] (z2) {};

    \draw [->] (delta) -- ++(-1.5,0) |- (w1) -- ++(0.8,0) node[midway] {$p$};
    \draw [<-] (P) -- ++(-2,0) node[left] {$w$};
    
    \draw [->] (z1) node[right, yshift=2mm ] {$q$} -- ++(0.75,0) |- (delta);
    \draw [->] (P) -- ++(2.,0) node[right] {$y$};

\end{tikzpicture}
\caption{LFT Representation of Nonlinear Systems.}
\flab{lft_basics}
\end{figure}
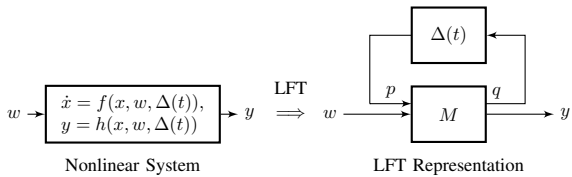

Consider a general nonlinear system of the form:
\begin{equation}
\dot{x} = f(x,w,\Delta(t)), \quad y = h(x,w,\Delta(t))
\end{equation}
where $x \in \mathbb{R}^n$ is the state, $w \in \mathbb{R}^m$ is the exogenous input, $y \in \mathbb{R}^p$ is the output, and $\Delta(t)$ represents time-varying nonlinear functions or uncertain parameters.

The LFT framework represents this system as:
\begin{equation}
G(\Delta(t)) = M_{22} + M_{21}\Delta(t)(I - M_{11}\Delta(t))^{-1}M_{12}
\elab{lft_general}
\end{equation}
where $M$ is a fixed LTI system partitioned as:
\begin{equation}
M = \begin{bmatrix} M_{11} & M_{12} \\ M_{21} & M_{22} \end{bmatrix}
\end{equation}

The linear system $M$ can be represented in state-space form as:
\begin{equation}
\begin{aligned}
\dot{x} &= Ax + B_1w + B_2p\\
q &= C_1x + D_{11}w + D_{12}p\\
y &= C_2x + D_{21}w + D_{22}p
\end{aligned}
\end{equation}

where $p = \Delta(t)q$ completes the feedback loop.

For rational (polynomial) nonlinear systems, the LFT representation can be exact, while general nonlinear systems require approximations. Common approximation methods include Taylor series expansions, Padé approximants for transcendental functions, sector-bounded approximations, and function substitution. Higher-order approximations improve accuracy but increase model complexity. The approximation error must be quantified to maintain valid performance guarantees.

\subsection{LFT Modeling of CR3B Dynamics}
The nonlinear equation of motion in \eqn{cr3bp_dynamics} can be expressed in state-space form as
\begin{subequations}
\begin{align}
\ddot{x} - 2\dot{y} - x &= -\frac{1 - \mu}{r_1^3}(x + \mu) - \frac{\mu}{r_2^3}(x - 1 + \mu), \\
\ddot{y} + 2\dot{x} - y &= -\frac{1 - \mu}{r_1^3}y - \frac{\mu}{r_2^3}y,\\
\ddot{z} &= -\left(\frac{1 - \mu}{r_1^3} + \frac{\mu}{r_2^3}\right)z,
\end{align}
\elab{normalized_dynamics}
\end{subequations}
where $r_1$ and $r_2$ are defined in \eqn{r1r2}.
The nonlinearities in \eqn{normalized_dynamics} are in $r_1(x,y,z)$ and  $r_2(x,y,z)$, which can be expressed as a parameter dependent system 

\begin{align}
\frac{d}{dt}\begin{pmatrix}x \\ y \\ z \\ \dot{x} \\ \dot{y} \\ \dot{z} \end{pmatrix} &=
\begin{bmatrix}
0_{3\times 3} & I_{3\times 3} \\
A_{21}(r_1,r_2) & A_{22}
\end{bmatrix}
\begin{pmatrix}x \\ y \\ z \\ \dot{x} \\ \dot{y} \\ \dot{z} \end{pmatrix} +\begin{pmatrix} 0 \\ 0 \\0 \\b_4(r_1,r_2)\\0 \\ 0
\end{pmatrix},
\elab{lft_dynamics}
\end{align}
where
\begin{align*}
A_{21}(r_1,r_2) &:= \textbf{diag}\left(a_{41}(r_1,r_2), a_{52}(r_1,r_2), a_{63}(r_1,r_2)\right), \\
A_{22} &:= \begin{bmatrix}
0 & 2 & 0\\
-2 & 0 & 0\\
0 & 0 & 0
\end{bmatrix},\\
a_{41}(r_1,r_2) &:= (\mu - 1)/r_1^3 - \mu/r_2^3 + 1,\\
a_{52}(r_1,r_2) &:= (\mu - 1)/r_1^3 - \mu/r_2^3 + 1,\\
a_{63}(r_1,r_2) &:= (\mu - 1)/r_1^3 - \mu/r_2^3,\\
b_4(r_1,r_2) &:= \mu(1-\mu)(1/r_2^3-1/r_1^3).
\end{align*}

To apply structured uncertainty modeling and LFT representation to the nonlinear CR3B dynamics, we must establish finite bounds on the distance functions $r_1$ and $r_2$ that encompass all physically realizable trajectory states within the operational envelope. The bounds on $r_1$ and $r_2$ can be established through analysis of the NRHO geometry, leveraging known periapsis and apoapsis distances, as well as maximum out-of-plane excursions. Additionally, numerical simulations of representative NRHO trajectories can be employed to empirically determine the extremal values of $r_1$ and $r_2$ over multiple orbital periods. These bounds should account for perturbations and maneuver-induced deviations to ensure robustness.

In general, assuming $r_1\in[r_{1_\text{min}},r_{1_\text{max}}]$, and $r_2\in[r_{2_\text{min}},r_{2_\text{max}}]$, we can treat them as multiplicative structured uncertainty, i.e.,
$$
r_1 = \Bar{r}_1(1+\delta_1\tilde{r}_1), \text{ and } r_2 = \Bar{r}_2(1+\delta_2\tilde{r}_2),
$$
where $\delta_1 \in [-1,1]$ and $\delta_2 \in [-1,1]$, and $\Bar{r}_1$ and $\bar{r}_2$ are nominal values, often taken as the average of the extreme values. 

All perturbation effects -- including SRP parameter variations, third-body position uncertainties, gravitational harmonic coefficient errors, and maneuver execution dispersions -- can be systematically captured as state-dependent norm-bounded disturbances. Each perturbation source is characterized by its maximum deviation from nominal values, with bounds that may vary with orbital phase, celestial body proximity, or operational mode. The resulting uncertain parameters $\{\delta_{w_i}\}$ satisfy $\|\delta_{w_i}\|_\infty \leq 1$. The exogenous disturbance inputs $w(t) = W_w(\delta_{w_1}, \delta_{w_2}, \ldots)\Bar{w}(t)$, with $\|\Bar{w}(t)\|_2 = 1$, aggregate these effects. This formulation allows the navigation filter to explicitly account for worst-case perturbation scenarios, ensuring robust performance across the full range of operational conditions.

We can express the nonlinear uncertain `parameters' in LFT form, using MATLAB's Robust Control Toolbox~\cite{matlabRobustControlToolbox}, which provides functions for LFT modeling and synthesis.

\subsection{LFT Models of Onboard Sensing System}
Drawing inspiration from the Orion optical navigation (OpNav) system deployed on Artemis 1 \cite{inman2024artemis, inman2024demonstration, christian2012onboard}, we consider the development of a celestial navigation system utilizing bearing-only measurements to Earth and Moon. We assume the optical sensor images Earth and Moon against the stellar background, extracting bearing angles through centroiding or limb-fitting techniques, and provides line-of-sight direction vectors from the spacecraft to each celestial body. The unit vectors are defined as
\begin{subequations}
\begin{align}
\uvec{e}_1 &= -\frac{\vo{r}_1}{r_1} = -\frac{1}{r_1}\begin{pmatrix} x + \mu \\ y \\ z \end{pmatrix}, \\
\uvec{e}_2 &= -\frac{\vo{r}_2}{r_2} = -\frac{1}{r_2}\begin{pmatrix} x - (1-\mu) \\ y \\ z \end{pmatrix},
\end{align}
\elab{unit_vectors}
\end{subequations}
where $\vo{r}_1$ and $\vo{r}_2$ are the position vectors from Earth and Moon to the spacecraft, respectively.

The measurement outputs are the components of these unit vectors, and can be compactly expressed as
\begin{equation}
y_m = -\begin{bmatrix} I_3/r_1 & 0_{3\times 3} \\ I_3/r_2 & 0_{3\times 3} \end{bmatrix} 
\begin{pmatrix}x \\ y \\ z \\ \dot{x} \\ \dot{y} \\ \dot{z} \end{pmatrix}  +  \begin{pmatrix} -\mu/r_1 \\ 0 \\ 0 \\ (1-\mu)/r_2 \\ 0 \\ 0 \end{pmatrix} + v,
\elab{sensor}
\end{equation}
where $v$ is the measurement noise vector. The sensor model in \eqn{sensor} is nonlinear in $r_1$ and $r_2$, which can be expressed in LFT form using the same approach as the dynamics model.

In optical navigation systems, sensor measurement noise $v$ exhibits strong dependence on target range due to signal-to-noise ratio degradation with increasing distance. As apparent target brightness decreases, the precision of image-based measurements -- such as centroids or limb features -- deteriorates correspondingly. This effect becomes particularly pronounced in the cislunar environment, where large and varying Earth-Moon distances amplify photon-limited sensing impacts. To model this phenomenon within a robust estimation framework, we express sensor noise as range-weighted. Consequently, the measurement noise in \eqn{sensor} can be expressed as
\begin{equation}
v = \textbf{blkdiag}\left(W_1(r_1)I_3,W_2(r_2)I_3\right)\bar{v},    
\elab{noise_model}
\end{equation}
where $\bar{v}$ is a unit-norm exogenous noise, and $W_k(\cdot)$ represents a range-dependent weighting function capturing sensor noise growth with distance. A physically motivated choice is $W_k(r_k) = \sqrt{\alpha_k} r_k$, yielding quadratic noise energy scaling with range: $\|(W_k(r_k)\bar{v_k})\|^2 = \alpha_k r_k^2$, where $r_k$ is either $r_1$ or $r_2$ depending on the sensor. Such representation is validated by optical navigation data from missions like Artemis I, where increased range to Earth or Moon resulted in visibly higher measurement uncertainty \cite{inman2024artemis}. 

The nonlinear measurement model in \eqn{sensor}, combined with the state-dependent noise modeled in \eqn{noise_model}, can be expressed in LFT form using MATLAB's Robust Control Toolbox~\cite{matlabRobustControlToolbox}.

\subsection{Synthesis of Robust $\mathcal{H}_\infty$ Observer}
With the nonlinear CR3BP dynamics in \eqn{lft_dynamics} and the sensor model in \eqn{sensor}, we can express the complete system in a linear parameter-varying (LPV) framework. This representation characterizes the system through explicit dependence on the nonlinear terms $r_1$ and $r_2$ as time-varying parameters. The resulting state-space formulation of this parameterized system can be written as

\begin{equation}
\begin{split}
\dot{x}_s &= A(\rho)x_s + B_ww + b(\rho),\\
y_m &= C_y(\rho)x_s + D_w(\rho)w + d(\rho),\\
z &= C_zx_s,
\end{split}
\elab{lft_system}
\end{equation}
where \(x_s\) denotes the system state vector, \(y_m\) represents the vector of measurements, and \(w\) encompasses all exogenous inputs, such as process disturbances and measurement noise. The vector $z$ indicates the quantities of interest to be estimated, which in our navigation context corresponds to the position components $(x,y,z)$. The parameter vector $\rho$ encapsulates the nonlinear elements $r_1$ and $r_2$, formally defined as $\rho := \begin{pmatrix}r_1 & r_2\end{pmatrix}^T$. Each of the system operators \(A(\rho)\), \(b(\rho)\), \(C_y(\rho)\), \(D_w(\rho)\), and $d(\rho)$ exhibits nonlinear dependence on $\rho$. By formulating these nonlinearities as structured uncertainties within the LFT framework, we can leverage robust estimation methodologies for observer synthesis.

The objective is to design a state estimator that provides accurate estimates of the outputs of interest \(z\) based on the measurements \(y_m\), while ensuring robustness against uncertainties in the parameters \(\rho\) and exogenous inputs \(w\). The schematic of the estimator is shown in \fig{Block Diagram}. The estimator can be designed using the $\mathcal{H}_\infty$ and $\mu$ synthesis framework, which enables the incorporation of structured uncertainties into the system.

\begin{figure}
\centering
\begin{tikzpicture}[auto, node distance=2cm, >=latex', thick, scale=0.9, every node/.style={scale=0.9}]
    \node [block, right of=d, node distance=2cm] (P) at (2,-0.25){Plant};
    \node [sum, right of=P, node distance=2cm, thick] (sum1) {\tiny $+$};
    \node [noise, above of=sum1, node distance=1cm] (noise) {};
    \node [block, right of=sum1, node distance=2cm] (observer) {Observer};
    \node [output, right of=observer] (zh) {};
    \node [output1, above of=zh, node distance=1.5cm] (z) {};
    
    \node [input, left of=P, yshift=0mm] (d) {$d$};
    \node [input, left of=P, yshift=-2mm] (u) {$u$};
    \coordinate [left of=P, yshift=-2mm, xshift=0.6cm] (u1) {};
    \coordinate [below of=u1, node distance = 10mm] (u2) {};
    
    \draw [->] (P) -- node {$y$} (sum1);
    \draw [->] (noise) -- node {$n$} (sum1);
    \draw [->] (sum1) -- node {$y_m$} (observer);
    \draw [->] (observer) -- (zh) node [at end, above, above left] {$\hat{z}$};
    \draw [->] (P) |- (z) node [at end, above left] {$z$};
    \draw [->] (d) node[left] {$d$} -- ++(1.29,0);

\end{tikzpicture}
\caption{System interconnection for designing and implementing proposed estimators.}
\flab{Block Diagram}
\end{figure}
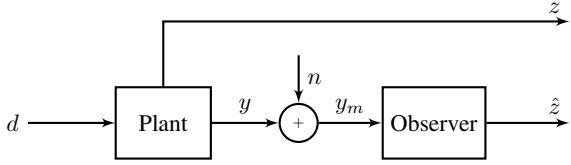

Following the derivation of standard full-order $\mathcal{H}_{\infty}$ observer \cite{duan2013lmis} (pg. 293), we propose the following observer structure for the system in \eqn{lft_system},
\begin{equation}
\dot{\hat{x}} = \big(A(\rho) + LC_y(\rho)\big)\hat{x} - L\big(y_m-d(\rho)\big) + b(\rho),
\elab{obs}
\end{equation}
which is slightly different from the standard observer structure due to the presence of the terms $b(\rho)$ and $d(\rho)$ in \eqn{lft_system}. In LPV models, the parameter $\rho$ is assumed to be known at runtime -- either via direct measurement or estimation.

Defining the error as $e(t) := x_s(t) - \hat{x}(t)$, we get the following error dynamics
\begin{align*}
\dot{e} = \big(A(\rho) + LC_y(\rho)\big)e + \big(B_w + LD_w(\rho)\big)w,
\end{align*}
which is in the form shown in Duan \etal \cite{duan2013lmis} (pg. 293). The estimated quantity of interest is $\hat{z} := C_z\hat{x}$, and the error in the estimate $\tilde{z}$ is related to the error $e$ as $\tilde{z} = C_ze$. Therefore, the dynamical system that relates exogenous input $w(t)$ to estimation error $\tilde{z}(t)$ is given by
\begin{equation}
\begin{split}
\dot{e} &= \big(A(\rho) + LC_y(\rho)\big)e + \big(B_w + LD_w(\rho)\big)w,\\
\tilde{z} &= C_ze.
\end{split}
\elab{estim-error}
\end{equation}
The objective is to determine $L$ such that $\|\tilde{z}(t)\|_2$ is minimized, which is achieved by minimizing the $\mathcal{H}_\infty$ norm of the system in \eqn{estim-error} \cite{doyle2013feedback}.

To synthesize the observer gain $L$, we can express the system in LFT form. The LFT representation of the system in \eqn{estim-error} can be written as
\begin{equation}
\begin{split}
  \dot{e} &= A(\rho)e + B_ww + \tilde{u}, \\
  \tilde{y}_m &= C_y(\rho)e + D_w(\rho) w,\\
  \tilde{u} &= L\tilde{y}_m,\\
  \tilde{z} &= C_ze.
\end{split}
\elab{lft_estim}
\end{equation}

The system in \eqn{lft_estim} can be expressed in the LFT form as shown in \fig{lft_estim}.
\begin{figure}
  \centering
\begin{tikzpicture}[auto, node distance=1.5cm, >=latex']
    
    \node [block] (P) at (0,0){$M$};
    \node [block, above of=P] (delta) {$\textbf{blkdiag}\left(\rho_1 I_{n_1},\rho_2 I_{n_2}\right)$};
    \node [block, below of=P] (L) {$L$};

    \node [input, left of=P, yshift=2mm] (w1) {};
    \node [input, left of=P, yshift=0mm] (w2) {};
    \node [input, left of=P, yshift=-2mm] (u) {};

    \node [output, right of=P, xshift=-.8cm, yshift=2mm] (z1) {};
    \node [output, right of=P, yshift=0mm] (z2) {};
    \node [output, right of=P, xshift=-.8cm, yshift=-2mm] (ym) {};

    \draw [->] (delta) -- ++(-2.25,0) |- (w1) -- ++(0.8,0);
    \draw [<-] (P) -- ++(-2.5,0) node[left] {$w$};
    \draw [->] (L) -| node[left] {$\tilde{u}$} (u) -- ++(0.8,0);

    \draw [->] (z1) -- ++(1.5,0) |- (delta);
    \draw [->] (ym) -- ++(0.8,0) |- node [midway,right] {$\tilde{y}_m$}(L);
    \draw [->] (P) -- ++(2.5,0) node[right] {$\tilde{z}$};

\end{tikzpicture}
\caption{LFT interconnection for designing robust $\mathcal{H}_\infty$ estimator.}
\flab{lft_estim}
\end{figure}
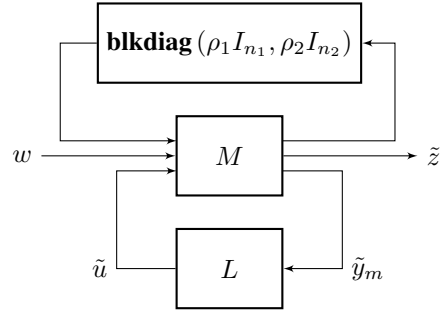

The robust observer synthesis involves parameterizing the uncertain dynamics using MATLAB's Robust Control Toolbox \cite{matlabRobustControlToolbox}. Specifically, we represent the parameters $r_1$ and $r_2$ as \texttt{ureal} objects with appropriate bounds, constructing the structured uncertainty blocks shown in the LFT interconnection of \fig{lft_estim}. For the robust estimator design, we apply \texttt{hinfstruct($\cdot$)} to obtain the optimal $\mathcal{H}_\infty$ observer $L^\ast$.

\section{Simulation Results}
\subsection{Simulation Setup}
We evaluate our approach using a simulation scenario where a spacecraft in NRHO estimates its position using bearing measurements to Earth and Moon. The simulation begins with the following initial conditions,
\begin{align}
x(0) := \begin{pmatrix}
    1.02950089 \\ 0 \\ -0.18680810 \\ 0 \\ -0.11898000 \\ 0\end{pmatrix},
\tilde{x}(0) := \begin{pmatrix}
    0.26\\
   -0.13\\
    0.13\\
    0.68\\
   -0.29\\
    0.29 \end{pmatrix}\hspace{-1mm}\times  10^{-4}.
\end{align}

The perturbation is typical of an NRHO trajectory, with a period of approximately 6.5 days. The values are representative of typical navigation uncertainties encountered after initial orbit determination for an $L_2$ southern NRHO at a $y=0$ crossing. The error profile follows established patterns observed in deep space navigation: largest uncertainties in the along-track direction, moderate cross-track errors, and smaller radial components, with velocity uncertainties in the centimeters-per-second range. The initial state estimate $\hat{x}(0)$ is set as $\hat{x}(0) = x(0) - \tilde{x}(0)$. The bounds for parameters $\rho$ are derived from trajectory analysis and summarized in Table \ref{tab:rho_bounds}.

\begin{table}
\begin{center}
\begin{tabular}{l||r|r|}
    & $\rho_1:=r_1$ & $\rho_2:=r_2$ \\[1mm] \hline
Min & 0.9495 &  0.0111\\
Max & 1.1112 & 0.2010\\
\end{tabular}
\caption{Bounds for parameters.}
\label{tab:rho_bounds}
\end{center}
\end{table}

The exogenous input vector $w(t)$ incorporates both process and measurement disturbances:
$$w(t) = \begin{pmatrix} d^T(t) & n^T(t) \end{pmatrix}^T,$$ where
$d(t) := \begin{pmatrix}d_x(t) & d_y(t) & d_z(t)\end{pmatrix}^T$, and $n(t) := \begin{pmatrix}n_1(t) & n_2(t) & n_3(t) & n_4(t) & n_5(t) & n_6(t)\end{pmatrix}^T.$ Here, $d_x(t)$, $d_y(t)$, and $d_z(t)$ represent perturbation accelerations (from sources including solar gravity, Jupiter's influence, and Earth's gravitational harmonics) in the $x$, $y$, and $z$ directions respectively. The terms $n_1(t)$ through $n_6(t)$ model sensor measurement noise. 

The corresponding input matrices are
\begin{align*}
B_w(\rho) &:= \begin{bmatrix}
0_{3\times 3} & 0_{3\times 6}\\ I_3 & 0_{3\times 6}
\end{bmatrix},\\ 
D_w(\rho) &:= \begin{bmatrix}
0_{6\times 3} & \textbf{blkdiag}\left(W_1(r_1)I_3,W_2(r_2)I_3\right)
\end{bmatrix},
\end{align*}
where the range-dependent weighting functions are defined as,
\begin{equation}
\begin{split}
W_1(r_1) &:= \eta_\text{min} + \frac{r_1 - r_{1_\text{min}}}{r_{1_\text{max}} - r_{1_\text{min}}}(\eta_\text{max} - \eta_\text{min}),\\
W_2(r_2) &:= \eta_\text{min} + \frac{r_2 - r_{2_\text{min}}}{r_{2_\text{max}} - r_{2_\text{min}}}(\eta_\text{max} - \eta_\text{min}).
\end{split}
\elab{sensor-noises}
\end{equation}

Following realistic optical navigation characteristics \cite{wu2022autonomous}, we set $\eta_\text{max} = 500$ arcsec and $\eta_\text{min} = 50$ arcsec, providing a noise profile that scales with target distance.  The optical sensors are assumed to be star-tracker quality, with noise modeled as band-limited white noise with a cutoff frequency of 0.1 Hz. Process noise is modeled as uniform white noise in the $[-0.01, 0.01]$ range, a relatively large disturbance in normalized CR3BP dynamics, deliberately chosen to stress-test observer robustness.

Since our primary interest lies in position estimation, we define the output matrix as $C_z := \begin{bmatrix}I_3 & 0_{3\times 3}\end{bmatrix}$. The remaining matrices in \eqn{lft_estim} are as defined in \eqn{lft_dynamics} and \eqn{sensor}, with $I_n$ denoting the $n\times n$ identity matrix.

\subsection{Computation of $r_1$ and $r_2$ from Line-of-Sight Vectors}
In the normalized rotating frame of the Earth-Moon Circular Restricted Three-Body Problem (CR3BP),
the Earth and Moon are located at fixed positions
\[
E = (-\mu,0,0)^\top, \qquad
M = (1-\mu,0,0)^\top.
\]

The Earth--Moon baseline is normalized to unity, i.e., $D = \| M - E \| = 1$. Suppose a spacecraft at position $S$ has access to the line-of-sight unit vectors
\[
\uvec{e}_1 = \frac{E-S}{\|E-S\|}, \qquad
\uvec{e}_2 = \frac{M-S}{\|M-S\|},
\]
pointing from the spacecraft to the Earth and Moon, respectively. Then $r_1 = \|S-E\|$ and $r_2 = \|S-M\|$. The vector closure condition for the Earth--Moon baseline is
\begin{equation}
M - E = r_1 \uvec{e}_1 - r_2 \uvec{e}_2.
\label{eq:closure}
\end{equation}
Since $M - E = D \, \uvec{e}_x$ with $\uvec{e}_x = (1,0,0)^\top$ in the rotating frame,
\eqref{eq:closure} becomes
\begin{equation}
D \uvec{e}_x = r_1 \uvec{e}_1 - r_2 \uvec{e}_2.
\end{equation}

Taking dot products with $\uvec{e}_1$ and $\uvec{e}_2$ yields two linear equations in $r_1,r_2$:
\begin{align}
D \,(\uvec{e}_x\cdot \uvec{e}_1) &= r_1 - (\uvec{e}_1 \cdot \uvec{e}_2)\, r_2, \\
D \,(\uvec{e}_x \cdot \uvec{e}_2) &= (\uvec{e}_1 \cdot \uvec{e}_2)\, r_1 - r_2.
\end{align}
Define the scalars
\[
c = \uvec{e}_1\cdot\uvec{e}_2,\, \alpha = \uvec{e}_1 \cdot \uvec{e}_x,\, \beta = \uvec{e}_2 \cdot \uvec{e}_x.
\]

The above system has the closed-form solution
\begin{equation}
r_1 = \frac{D(c\beta-\alpha)}{1-c^2}, \qquad
r_2 = \frac{D(\beta - c\alpha)}{1-c^2}.
\label{eq:range-formula}
\end{equation}
Since $D=1$ in normalized units, the ranges are dimensionless.

The geometry is well-conditioned provided $1-c^2$ is not too small, i.e.\ the spacecraft, Earth, and Moon are not nearly collinear. A residual check for closure,
\[
r = \left\| D \uvec{e}_x - (r_1 \uvec{e}_1 - r_2 \uvec{e}_2) \right\|,
\]
should be close to zero in a consistent solution. Additionally, the ranges $r_1,r_2$ must be positive.

Equation \eqref{eq:range-formula} provides a method to compute $r_1$ and $r_2$ from line-of-sight measurements. When $1-c^2$ is small (near-collinear geometry), the system becomes ill-conditioned. In such cases, $r_1$ and $r_2$ can be computed directly from state estimates $\hat{x}(t)$, assuming small estimation errors.

In NRHOs, the Earth, Moon, and spacecraft never become collinear, maintaining an angle that keeps the denominator $1 - c^2 = \sin^2(\angle ESM)$ bounded away from zero. This ensures well-conditioned range computations throughout the orbit. In our simulations, we compute $r_1$ and $r_2$ directly from the noisy bearing measurements $y_m$, reflecting realistic onboard navigation scenarios while testing estimator robustness against both measurement noise and nonlinearities.

\subsection{Observer Implementation and Results}
\fig{errors} shows the estimation errors in position and velocity 
components over three nondimensional time units (TU) for the proposed robust estimator. 
The position errors (top row) remain well below $10^{-5}$ in normalized distance units 
($\sim 4$~km equivalent), with the largest deviations occurring in the out-of-plane coordinate($z$), where the NRHO dynamics are most stiff. These errors are bounded and exhibit smooth 
behavior across the trajectory, indicating that the estimator is able to track the nominal 
state effectively even during periods of strong out-of-plane motion.

The velocity errors (bottom row) exhibit sharper features due to the time-varying stiffness 
of NRHO dynamics, particularly near perilune passages where restoring forces change rapidly. 
The largest errors occur in the out-of-plane velocity $\dot{z}$, consistent with 
CR3BP vertical dynamics where small deviations in $z$ lead to rapid changes in $\dot{z}$. 
Additionally, line-of-sight measurements provide less information in the vertical direction, 
reducing $\dot{z}$ observability. 

In summary, the robust estimator maintains bounded errors across all states, with smooth 
position tracking throughout the orbit and short-duration excursions in the velocity errors 
near perilune.

\begin{figure}
    \centering
    \includegraphics[width=0.48\textwidth]{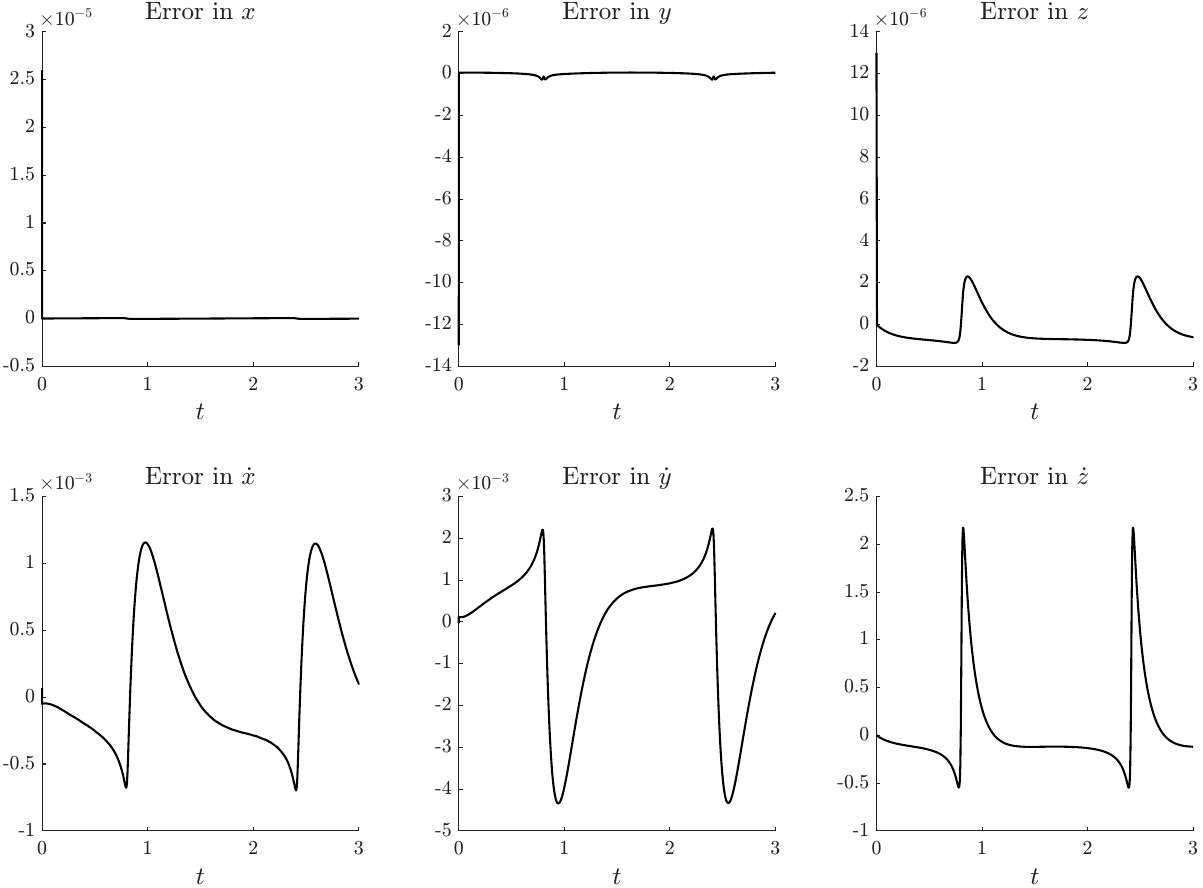}
    \caption{Estimation errors using the proposed robust $\mathcal{H}_\infty$ observer.}
    \flab{errors}   
\end{figure}

\section{Conclusions}
This paper has presented a robust autonomous navigation framework for cislunar operations 
based on Linear Fractional Transformation (LFT) modeling of the nonlinear CR3BP dynamics 
and range-dependent optical sensor uncertainty. A full-order $\mathcal{H}_\infty$ observer 
was synthesized to provide guaranteed $\mathcal{L}_2$ performance bounds. The proposed 
formulation embeds both the orbital dynamics and sensing characteristics in an LFT 
representation, enabling a nonlinear estimator that operates directly on the governing equations. This approach avoids the inaccuracies that can arise when relying on local 
linearizations of the strongly nonlinear CR3BP dynamics. By treating the nonlinearities as 
structured uncertainties and incorporating range-weighted sensor models, the framework 
systematically addresses worst-case geometric conditions and perturbation effects 
characteristic of NRHO environments. The bearing-only measurement strategy is aligned with 
practical spacecraft constraints, relying on passive, star-tracker--class optical sensors 
that provide continuous Earth and Moon line-of-sight observations without dependence on 
Deep Space Network support, high-precision timing references, or continuous ground contact.

The simulation results demonstrate that the proposed estimator maintains bounded errors and 
smooth position tracking over multiple orbital periods in the presence of NRHO dynamics. The 
largest deviations occur in the out-of-plane states, consistent with the reduced 
observability and increased stiffness of the vertical dynamics. These results indicate that 
the framework provides a viable approach to achieving guaranteed-performance navigation in 
cislunar space when using flight-representative sensors and disturbance models, and thereby 
supports the development of autonomous operations in this regime.

\section*{Acknowledgment}
This work is supported by AFOSR grant FA9550-22-1-0539 with Dr. Erik Blasch as the program director.

\bibliographystyle{IEEEtran}
\bibliography{refs}

\begin{thebibliography}{10}
\providecommand{\url}[1]{#1}
\csname url@samestyle\endcsname
\providecommand{\newblock}{\relax}
\providecommand{\bibinfo}[2]{#2}
\providecommand{\BIBentrySTDinterwordspacing}{\spaceskip=0pt\relax}
\providecommand{\BIBentryALTinterwordstretchfactor}{4}
\providecommand{\BIBentryALTinterwordspacing}{\spaceskip=\fontdimen2\font plus
\BIBentryALTinterwordstretchfactor\fontdimen3\font minus \fontdimen4\font\relax}
\providecommand{\BIBforeignlanguage}[2]{{%
\expandafter\ifx\csname l@#1\endcsname\relax
\typeout{** WARNING: IEEEtran.bst: No hyphenation pattern has been}%
\typeout{** loaded for the language `#1'. Using the pattern for}%
\typeout{** the default language instead.}%
\else
\language=\csname l@#1\endcsname
\fi
#2}}
\providecommand{\BIBdecl}{\relax}
\BIBdecl

\bibitem{montenbruck2000satellite}
O.~Montenbruck and E.~Gill, \emph{Satellite Orbits: Models, Methods, and Applications}.\hskip 1em plus 0.5em minus 0.4em\relax Springer, 2000.

\bibitem{carpenter2025navigation}
J.~R. Carpenter and C.~N. D'Souza, ``{Navigation Filter Best Practices. NASA/TP–2018–219822/Revision},'' Tech. Rep., 2025.

\bibitem{julier2004unscented}
S.~J. Julier and J.~K. Uhlmann, ``Unscented filtering and nonlinear estimation,'' \emph{Proceedings of the IEEE}, vol.~92, no.~3, pp. 401--422, 2004.

\bibitem{gordon1993novel}
N.~J. Gordon, D.~J. Salmond, and A.~F. Smith, ``Novel approach to nonlinear/non-gaussian bayesian state estimation,'' in \emph{IEE proceedings F (radar and signal processing)}, vol. 140, no.~2.\hskip 1em plus 0.5em minus 0.4em\relax IET, 1993, pp. 107--113.

\bibitem{iiyama2021autonomous}
K.~Iiyama and R.~Funase, ``Autonomous and decentralized orbit determination and clock offset estimation of lunar navigation satellites using gps signals and inter-satellite ranging,'' in \emph{Proceedings of the 34th International Technical Meeting of the Satellite Division of the Institute of Navigation (ION GNSS+ 2021)}.\hskip 1em plus 0.5em minus 0.4em\relax St. Louis, MO, USA: Institute of Navigation, Sep. 2021, pp. ---, extended Abstract.

\bibitem{bar2001estimation}
Y.~Bar-Shalom, X.~R. Li, and T.~Kirubarajan, \emph{Estimation with applications to tracking and navigation: theory algorithms and software}.\hskip 1em plus 0.5em minus 0.4em\relax John Wiley \& Sons, 2001.

\bibitem{kumar2025cislunar}
T.~Kumar and R.~Bhattacharya, ``$\mathcal{H}_\infty$ optimal navigation in the cislunar space with lft models,'' in \emph{IEEE International Conference on Multi-Sensor Fusion and Integration for Intelligent Systems}, 2025.

\bibitem{zhou1998essentials}
K.~Zhou and J.~C. Doyle, \emph{{Essentials of Robust Control}}.\hskip 1em plus 0.5em minus 0.4em\relax Prentice Hall Upper Saddle River, NJ, 1998, vol. 104.

\bibitem{matlabRobustControlToolbox}
\BIBentryALTinterwordspacing
{The MathWorks, Inc.}, \emph{{Robust Control Toolbox User's Guide}}, MathWorks, Natick, Massachusetts, USA, 2024, version R2024a. [Online]. Available: \url{https://www.mathworks.com/products/robust.html}
\BIBentrySTDinterwordspacing

\bibitem{inman2024artemis}
R.~Inman, G.~Holt, J.~Christian, K.~W. Smith, and C.~D'Souza, ``{Artemis-I Optical Navigation System Performance},'' in \emph{AIAA SCITECH 2024 Forum}, 2024, p. 0514.

\bibitem{inman2024demonstration}
R.~J. Inman, ``{Demonstration of the Orion Optical Navigation System on Artemis I},'' in \emph{4th Space Imaging Workshop}.\hskip 1em plus 0.5em minus 0.4em\relax Georgia Institute of Technology, 2024.

\bibitem{christian2012onboard}
J.~A. Christian and E.~G. Lightsey, ``{Onboard Image-Processing Algorithm for a Spacecraft Optical Navigation Sensor System},'' \emph{Journal of Spacecraft and Rockets}, vol.~49, no.~2, pp. 337--352, 2012.

\bibitem{duan2013lmis}
G.-R. Duan and H.-H. Yu, \emph{{LMIs in Control Systems: Analysis, Design and Applications}}.\hskip 1em plus 0.5em minus 0.4em\relax CRC Press, 2013.

\bibitem{doyle2013feedback}
J.~C. Doyle, B.~A. Francis, and A.~R. Tannenbaum, \emph{{Feedback Control Theory}}.\hskip 1em plus 0.5em minus 0.4em\relax Courier Corporation, 2013.

\bibitem{wu2022autonomous}
C.~X. Wu, P.~Machuca, L.~Felicetti, and J.-P. Sanchez, ``{Autonomous Optical Navigation for Small Spacecraft in Cislunar Space},'' 2022.

\end{thebibliography}

\end{document}